\documentclass{ws-ijmpd}
\usepackage{graphics}
\usepackage{psfrag}

\newcommand{\electron}{\ensuremath{\gamma}}
\newcommand{\hardphoton}{\ensuremath{\varepsilon}}
\newcommand{\softphoton}{\ensuremath{\omega}}

\begin{document}

\markboth{A. Eungwanichayapant and F.A. Aharonian}
{Very High Energy $\gamma$-rays from $e^\pm$ Pair Halos}

\title{Very High Energy $\gamma$-rays from $e^\pm$ Pair Halos}

\author{A. Eungwanichayapant}
\address{School of Science, Mae Fah Luang University, 333 Ta-Sud\\
Muang, Chiang Rai, 57100, Thailand\\
anant.e@gmail.com}

\author{F. Aharonian }
\address{Dublin Institute for Advanced Studies, 
31 Fitzwilliam Place, Dublin 2, Ireland and 
Max-Planck-Institut f\"ur Kernphysik, Saupfercheckweg 1\\ 
69117 Heidelberg,  Germany\\
Felix.Aharonian@mpi-hd.mpg.de}

\maketitle

\begin{history}
\received{Day Month Year}
\revised{Day Month Year}
\end{history}

\begin{abstract}
In this paper we study the formation of 
giant electrons-positron pair halos 
around the powerful high energy extragalactic sources. 
We investigate the dependence of 
radiation of pair halos, in particular  
the spectral and angular distributions
on the energy spectrum of the primary gamma-rays, the 
redshift of the source, and the flux of 
the extragalactic background light.
\end{abstract}


\section{Introduction}

Interactions of very high energy (VHE) 
gamma-rays with the diffuse Extragalactic Background 
light (EBL)\cite{Nikishov,Gould&Schreder1967}, as well as the 
the possible effects caused by propagation of secondary 
(pair-produced) electrons in the intergalactic 
medium (see e.g. Ref. \refcite{FAetal02}), 
make the high energy gamma-rays unique 
carries of cosmological information about the epochs of 
formation of galaxies and their evolution in the past (see e.g. 
Ref.~\refcite{Dwek,Malkan&Stecker1998,Primack_et_al2000,Kneiske_et_al2002,Francescini08,Primack08}). 
The method is based on the distinct features 
in the spectra of high energy gamma-rays arriving from 
distant extragalactic objects caused by interactions 
with EBL. 

A serious obstacle in practical realization 
of this method is our poor knowledge concerning
the primary (intrinsic) gamma-ray spectra produced 
in the source \cite{FA01}.
It is believed that the well coordinated observations  
of gamma-ray blazars in different energy bands
should allow gamma-ray astronomers to identify 
the radiation mechanisms, derive the principal  model parameters, and 
thus reconstruct the intrinsic gamma-ray spectra based on
spectral and temporal properties of 
blazars (see e.g.  Ref. \refcite{CopAh99}).
Even so, for reliable estimates of the intergalactic 
absorption effect, and ultimately for derivation of the flux 
and spectrum of EBL, one should take into account  
possible integral absorption of gamma-rays. 
The importance of the internal photon-photon 
absorption in blazars has been discussed by many authors,
in particular just after the discovery by COS B of 
the first gamma-ray blazar - 3C273 \cite{McBreen}.
The complex  behavior of internal photon-photon 
absorption may result in a rather "irregular" 
deformation of the spectrum of primary gamma-rays (from \textit{steepening} or even sharp cutoffs to \textit{hardening})\cite{FAKhanCost}. This, as well as 
the effects related to the redshift of blazrs\cite{Anita}, 
challenge  the feasibility of reliable  
derivation of EBL based on absorption features in the 
gamma-ray spectra of extragalactic objects. 
 
Moreover, strictly speaking, the intergalactic absorption 
features  contain information about the product 
of the diffuse extragalactic background radiation 
density $w_{\rm r}$ and the Hubble constant $H_0$. 
Fortunately, it is  possible to decouple   
$w_{\rm r}$ and  $H_0$ by studying the spectral and angular characteristics of VHE $\gamma$ radiation from hypothetical electron-positron pair halos surrounding  
powerful nonthermal extragalactic objects. 
These giant structures 
are unavoidably formed around any extragalactic 
VHE source  due to development of pair cascades 
initiated by  interactions of primary 
multi-TeV photons with  EBL\cite{Halo_original}. 

\section{Basic processes of formation of pair halos} \label{sec:Interactions}
For formation of electron-positron pair halos 
around extragalactic objects, the photon-photon
pair production (PP)
and inverse Compton scattering (IC) are two
principal processes which initiate and support 
the electromagnetic cascade development\cite{Halo_original}. 
The mean free pathlengths and the energy spectra of 
secondary particles produced at  
these interactions are  described below.

 \subsection{Pair production}
 For the pair production at a collision
 of a gamma-ray photon  of energy   \hardphoton\
 with a target photon of  energy \softphoton\ the following 
 threshold conditions is required:
 \begin{equation}
  2 (1-{\beta^{\prime}}^2)^{-1} =  \hardphoton\softphoton(1-\cos \theta),
 \end{equation}
where $\beta^\prime$ is a velocity of the electron (positron) 
$e^\pm$ in the CM frame
and $\theta$ is an angle between the momenta of two photons in lab frame
(below we will use the energy of $e^\pm$ pairs in units of the electron rest mass, $m_{\rm e} c^2$).

\begin{figure}[tb]
  \begin{center}
   \scalebox{0.6}{\includegraphics{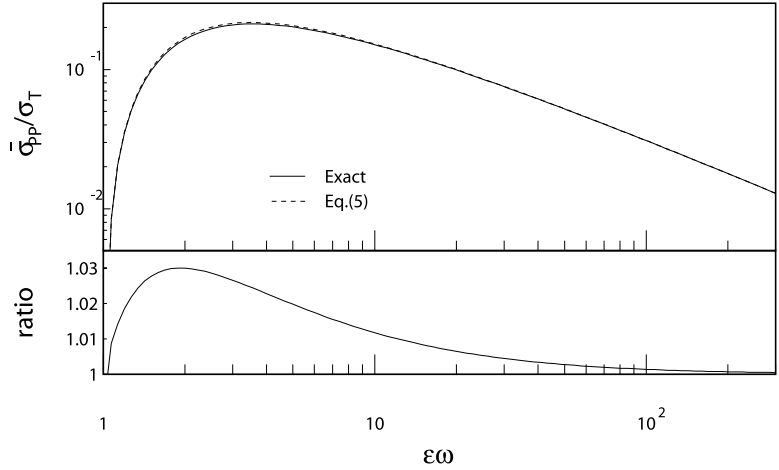}}
   \caption{Top panel: The cross section of photon-photon pair production
   $\overline{\sigma}_{PP}$
   (in units of the Thomson cross section $\sigma_T$)
calculated using the exact expression  (solid line) and the 
approximation given by 
Eq.(\ref{eq:pp-Aharonian} (dashed line). Bottom panel:
the ratio  of these two presentations of the 
cross-section.}
   \label{fig:pp-compare}
   \end{center}
\end{figure}

 The mean free path length $\Lambda_{PP}$ of a  gamma-ray of energy 
 \hardphoton\ traveling through the isotropically-distributed
 photon gas of  differential energy density $n(\softphoton)$ is
 \begin{equation} \label{eq:pp-int-cross}
 \Lambda_{PP}^{-1}(\hardphoton) = \int_{\softphoton_{th}}^{\infty}
                            \overline{\sigma}_{PP} n(\softphoton) d\softphoton,
 \end{equation}
 where $\overline{\sigma}$ is the cross-section 
 averaged over directions of interacting photons:
 \begin{equation} \label{eq:pp-ang-ave-cross}
   \overline{\sigma}_{PP} = \frac12 \int_{-1}^{1-2/s_0} (1-\mu)
                       \sigma_{PP} d\mu \ .
 \end{equation}
 Here $s_0 = \hardphoton\softphoton$, $\mu = \cos\theta$ and $\sigma_{PP}$ is
 the total cross section of the process  (see e.g.~Ref.~\refcite{Akhiezer&Berestetskii1965})
 \begin{eqnarray}
  \sigma_{PP} & = & \frac3{16} \sigma_T (1-{\beta^\prime}^2) {\beta^\prime} \left[
  \frac{(3-{\beta^\prime}^4)}{{\beta^\prime}} \ln \frac{1+{\beta^\prime}}{1-{\beta^\prime}} 
                \frac+ 2({\beta^\prime}^2-2) \right],
 \end{eqnarray}
 where $\sigma_T$ is the Thomson cross section.
 For integration of  Eq.(\ref{eq:pp-ang-ave-cross}) 
 we use the following approximation\cite{Aharonian2004}, 
 \begin{eqnarray} \label{eq:pp-Aharonian}
   \overline{\sigma}_{PP} & = & \frac{3 \sigma_T}{2s_0^2} \left[ \left( s_0
                           + \frac12 \ln s_0 - \frac16 + \frac1{2s_0} \right)\right.
                           \times \ln (\sqrt{s_0} + \sqrt{s_0-1})        \nonumber \\
                          &   & - \left. \left( s_0 + \frac49 - \frac1{9s_0}
                                \right) \sqrt{1 - \frac1{s_0}}
                                \right],
 \end{eqnarray}
 which provides an accuracy  better than 3\% 
 as demonstrated in Fig.\ref{fig:pp-compare}. 

\begin{figure}[tb]
   \begin{center}
       \scalebox{0.6}{\includegraphics{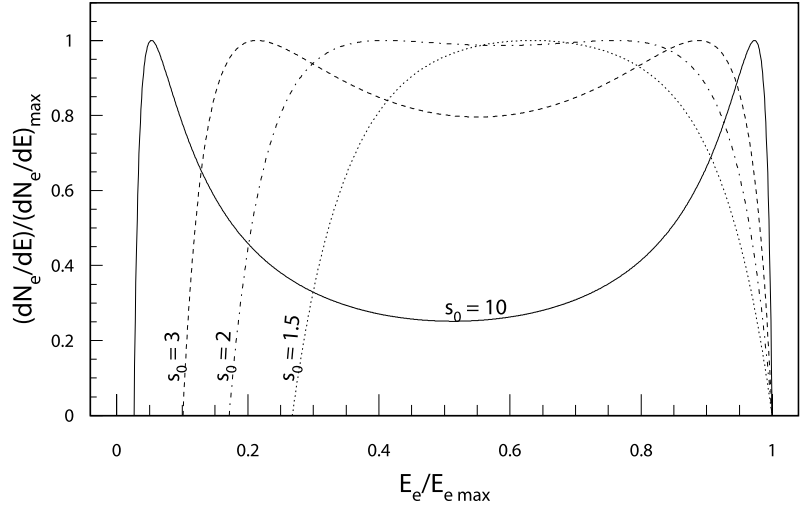}}
      \caption[Differential spectra of $e^\pm$ from PP]{\small{Differential spectra of 
      electrons and positrons  from photon-photon interactions 
      for different values of $s_0$:  10 (solid line),  3
 (dashed line),  2 (dot-dashed line) and 1.5 (dotted line).}}
 \label{fig:pp-diff-cross}
    \end{center}
 \end{figure}

 The resulting $e^\pm$ energy distribution is described as  
 \cite{Aharonian_et_al1983}
 \begin{eqnarray} \label{eq:pp_spectrum}
   \frac{dN_{e^\pm}(\electron , \softphoton, \hardphoton)}{d \electron}
   & = & \frac{3\sigma_T}{32\softphoton^2 \hardphoton^3}
         \left[ \frac{4\epsilon^2}{(\epsilon -\electron) \electron}\right.
         \times\ln\frac{4\softphoton(\epsilon-\electron)\electron}{\epsilon}
         -8\softphoton\epsilon \nonumber \\
   &   &  + \frac{2(2\softphoton\epsilon-1)\epsilon^2}{(\epsilon-\electron)\electron}
         - \left. \left( 1-\frac1{\softphoton\epsilon} \right)
\frac{\epsilon^4}{(\epsilon-\electron)^2\electron^2}
         \right],
 \end{eqnarray}
 where $\electron$ is the energy of the electron (positron) 
 (in $m_ec^2$ units) and $\epsilon = \hardphoton
 + \softphoton \approx \hardphoton$ is the photon energy in the
lab-frame.  The range of $\electron$ is
 \begin{equation}
   \frac{\epsilon}2\left( 1-\sqrt{1-\frac1{\epsilon \softphoton}}
   \right) \leq \electron \leq
   \frac{\epsilon}2\left( 1+\sqrt{1-\frac1{\epsilon \softphoton}}
   \right).
 \end{equation}
 Figure~\ref{fig:pp-diff-cross} shows the distributions of 
 electrons (positrons) for  different $s_0$.

 \subsection{Inverse Compton scattering}
 The mean free path of an electron due to the 
 inverse Compton scattering, $\Lambda_{IC}$, for the electron 
 propagating  through the isotropically 
 distributed photon gas density $n(w)$ is, 
 \begin{equation} \label{eq:ic-int-cross}
   \Lambda_{IC}^{-1} =
 \frac1{\beta}\int_0^{\infty} \overline{\sigma}_{IC}n(\softphoton) d\softphoton \ ,
 \end{equation}
where the averaged over the angles cross section is
 \begin{equation} \label{eq:ic-absorption-mod}
   \overline{\sigma}_{IC} = \frac12 \int_{-1}^{1} (1-\beta \mu)\sigma_{IC} d\mu \ .
 \end{equation}
 The total IC  cross-section in the lab frame (see e.g.
 Ref. \refcite{Akhiezer&Berestetskii1965}) can be presented in the form
 \begin{eqnarray}
  \sigma_{IC} & = & \frac{3 \sigma_T}{4\chi}
                    \left[ \left(1-\frac4{\chi}-\frac8{\chi^2}\right) \ln (1+\chi)
                    +\frac12 + \frac8{\chi} - \frac1{2(1+\chi)^2} \right],
 \end{eqnarray}
 where $\chi = 2w\electron(1- \beta \mu)$.
 The exact solution of Eq.(\ref{eq:ic-int-cross}) 
 can be presented in an analytical form 
 \cite{Aharonian_et_al1985,Protheroe1986}, 
 but the resulting expression contains a dilogarithm. A quite
 good approximation, with an accuracy better than 10\% 
 has been suggested in Ref.~\refcite{Coppi&Blandford1990}.

\begin{figure}[tb]
   \begin{center}
      \scalebox{0.6}{\includegraphics{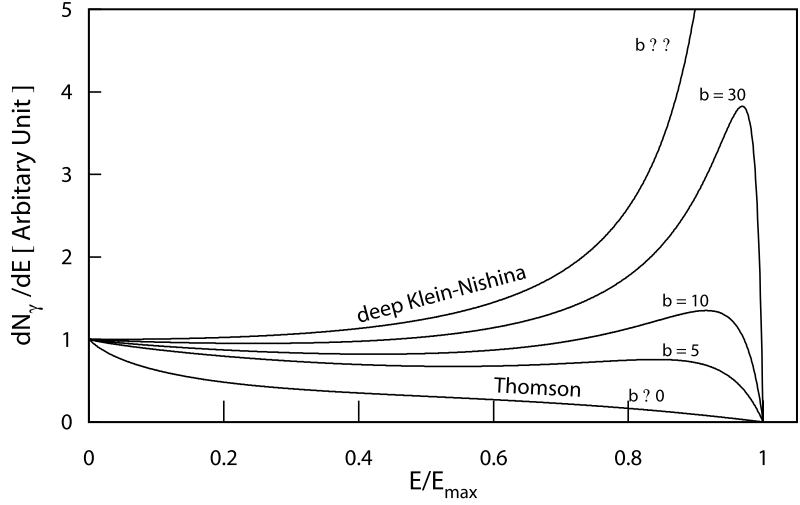}}
      \caption{\small{The  spectral energy
 distribution of upscattered photons (in arbitrary units) 
 for different values of the parameter $b$: $b\rightarrow 0,\
 b=5,\ 10,\  30,\ $ and $b\rightarrow\infty$ .}} \label{fig:ic_spectrum}
    \end{center}
 \end{figure}

 The spectral distribution of the scattered photons via IC for the
 isotropic photon gas is \cite{Blumenthal&Gould1970} 
 \begin{eqnarray} \label{eq:ic_spectrum}
   \frac{dN_{\hardphoton}}{d\hardphoton}
  & = & \frac{3\sigma_T}{4w \electron^2} \left[ 1+\frac{x^2}{2(1-x)}+\frac{x}{b(1-x)}
        -\frac{2x^2}{b^2(1-x)^2} \right.-\frac{x^3}{2b(1-x)^2}
        \nonumber \\
  &   & \left.-\frac{2x}{b(1-x)}\ln \frac{b(1-x)}x\right],
 \end{eqnarray}
 \begin{equation} \label{eq:ic energy limit}
   \frac{\softphoton}{\electron} \ll x \leq \frac{b}{1+b} =
   \frac{{\hardphoton}_{max}}{\electron}; \
  x \equiv \frac{\hardphoton}{\electron}; \
    b\equiv4\softphoton\electron.
 \end{equation}
 The spectra of upscattered photons depend only on the parameter
 $b$, as illustrated in Fig.~\ref{fig:ic_spectrum}. 
 If the value of $b$ is small, as in the Thomson regime, the upscattered photon population is 
 predominantly in the low energy regime. If the value of $b$ is large, corresponding to the 
 Klein-Nishina regime, the population is shifted towards high energies.

 \section{The method of calculations} \label{sec:method}

In order to  study of  formation of pair halos and  to calculate the 
characteristics of  their radiation, in this work  we  used the method 
of Monte Carlo simulations of interactions high energy electrons and photons 
with  intergalactic radiation fields.   This method provides an effective  tool for
study of development of electromagnetic cascades, and, in particular,  
allows  precise calculations of both energy and spatial distributions of  cascade  
particles for an arbitrary  background photon field. The method of calculations 
presented below is described in  Ref. \refcite{PhD}.   For  the treatment   of interaction lengths and  
energy distributions of electrons and photons produced at the photon-photon and electron-photon interactions
we used the corresponding total and differential cross-sections of these processes.  The 
hit-or-miss method\cite{James1980} and the rejection method\cite{Press_et_al1994} 
have been used for simulations of the  interaction lengths  and the energy distributions 
of cascade particles.
We assumed that  
the magnetic field in the intergalactic medium is sufficiently small, so the synchrotron cooling of electrons can be neglected.
On the other hand, we assumed that  the  magnetic field is sufficiently large, thus the secondary  electrons  of all relevant energies 
are isotropised  before  they interact with ambient photons.  These two conditions are safely satisfied 
for a very broad  range of the  intergalactic magnetic fields  between $10^{-7}$~G to $10^{-12}$~G.

The background photon fields used in our calculation consist of  
Cosmic Microwave Background (CMB) and the Extragalactic Background Light (EBL). 
Both the CMB and EBL are functions of redshift.   Since the background photon
field at  large cosmological redshifts  significantly differ from that at the present epoch, 
the assumption of a consistent background during the development of the cascade
cannot be held  for the large  redshift sources.
On the other hand, the cascade simulation with evolving background photon field are computationally expensive.
To make calculations faster, the simulation scheme was adjusted by simulating the
cascade only at a  certain epoch  or in the limited region around a source of redshift $z_s$, 
when  one can assume constant background radiation 
with the EBL flux corresponding  the  cosmological epoch of $z=z_s$. Obviously, 
the simulation region should be big enough to guarantee an adequate accuracy of calculations 
of the spatial distributions of particles around  the central source.  At the same time the 
region should be relatively compact so the constant background approximation can be applied. 
For example,  for a source at a distance of 1 Gpc, the gamma-ray 
production region around the source with an angular size $5^\circ$ 
corresponds to the linear size  $\approx 100$~Mpc, 
so we can safely assume that the EBL is not noticeably  
changed across this size or over the corresponding 
time  of $\leq 3 \times 10^{8}$~yr.   

\begin{figure}[tb]
\begin{center}
 \scalebox{0.6}{\includegraphics{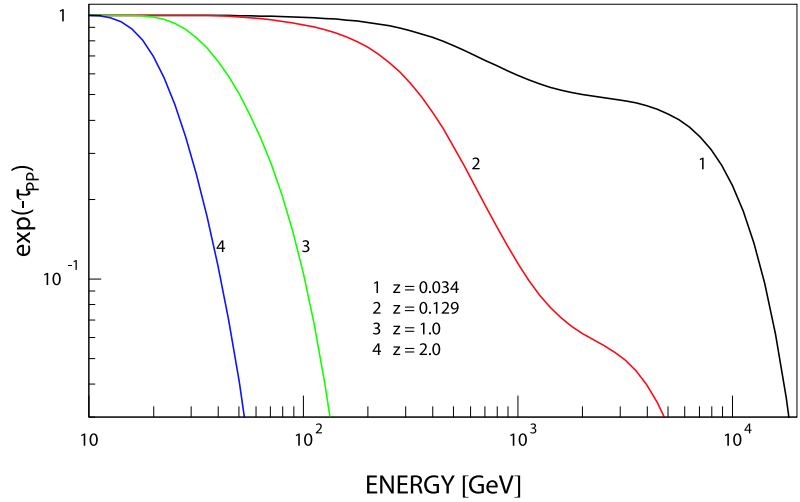}}
 \caption{\small{The attenuation factor $\exp(-\tau_{PP})$
 for gamma photons emitted from sources at different redshifts as a function of  energy for 
a specific EBL  proposed by Primack et al. (2000).}}
 \label{fig:attenuation factor}
\end{center}
\end{figure}

Beyond the "cascade  region",   we  take into account the interactions with EBL 
only in terms of absorption (i.e. without of the cascade treatment). Thus  the results presented 
in this paper are relevant only  for  the  core of halos within angular size of  a few degree. 
Although  pair halos do extend beyond $5^\circ$, their cores, in fact, 
present the most interest,  at least from the point  view of detectability of  these huge structures.
Therefore in most of cases, when we are interested in the compact cores of halos, 
the calculations are done using this approximate procedure.
Within this approach, to  calculate  the observed energy distribution  $dN/dE$,  
we modify the distribution $dN_{z_s}/dE$ calculated in the "cascade region"  
at  the epoch $z_s$,
by multiplying to the  attenuation factor, $\exp[-\tau_{PP}(\hardphoton)]$:
 \begin{equation}
  \frac{dN(\hardphoton)}{dE} = \frac{dN_{z_s}(\hardphoton)}{dE}
                               \exp[-\tau_{PP}(\hardphoton)],
 \end{equation}
 where $\tau_{PP}$ is the PP optical depth, 
 \begin{eqnarray} \label{eq:pp tau}
   \tau_{PP}(z_s,\hardphoton) & = &
   \frac{c}{H_0}\int_0^{z_s}dz(1+z)^{-2}
   \int_{1/{\hardphoton}_z}^{\infty}d\softphoton_z
   n(\softphoton_z,z)\overline{\sigma}_{PP}({\hardphoton}_z,\softphoton_z),
 \end{eqnarray}
 Here ${\hardphoton}_z$ and $\softphoton_z$ are the $\gamma$-ray and background photon
 energies at  the redshift $z$,  $H_0$ is the Hubble constant;  in this paper  we use $H_0$= 65 km/s/Mpc. 
 The attenuation factors at different
 redshifts are shown in Fig.~\ref{fig:attenuation factor}.

\section{Results}
Below we present results of calculations  of the  
spectral energy distributions (SEDs) and angular  distributions of 
gamma-rays from pair halos around sources located at different redshifts.  
The main uncertainty of predicted energy spectra and  morphology of pair 
halos  is related to the uncertainties of EBL, especially in the mid and far infrared wavelengths. 
Here we  use   three  different  EBL models proposed by  
Malkan \& Stecker (1998, hereafter M98)\cite{Malkan&Stecker1998}, 
Primack et al. (2000, hereafter P00)\cite{Primack_et_al2000} and 
Kneiske et al. (2002, hereafter K02)\cite{Kneiske_et_al2002}. 
One should  note the  recent developments related to the EBL at near infrared and 
optical wavelengths (see e.g. Ref. \refcite{Francescini08,Primack08}). However 
this wavelength  band is not critical for formation of pair halos. On the other hand the above  models represent quite broadly the possible range 
of EBL  at most principal  (from mid- to  far- infrared)
bands responsible for formation of pair halos around sources with redshifts up to $z \sim 2$. 

\begin{figure}[tb]
\begin{center}
  \scalebox{0.35}{\includegraphics{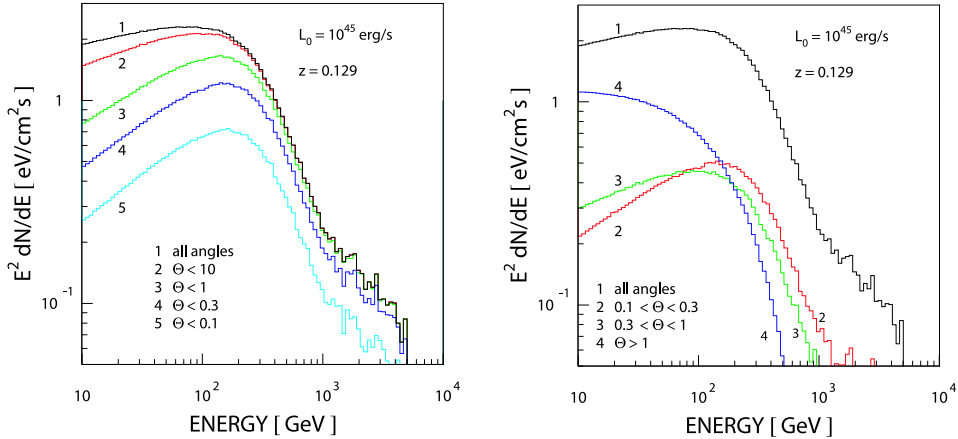}}
  \hspace{0.2cm}
  \caption{\small{ The SED of a pair halo  at $z=0.129$  
  within $\Theta \leq 10^\circ, \ 1^\circ, \ 
  0.3^\circ, \ \rm and \ 0.1^\circ$ (left panel) and $0.1^\circ \leq \Theta \leq 0.3^\circ$,  $0.3^\circ \leq \Theta \leq 1^\circ$, \ \rm and \ 
  $\Theta \geq 1^\circ$ (right panel).  In both panels the SEDs
   integrated over all angles are  also shown.}}  
  \label{fig:preliminary result2a}
\end{center}
\end{figure}

In Fig.~\ref{fig:preliminary result2a} we present the Spectral Energy Distribution (SED) of 
a pair halo within different angles formed around a source with redshift $z_s=0.129$. This 
corresponds to the redshift of the  blazar  1ES 1426+482, but  is  also representative for 
many  TeV blazars  reported  by the HESS, MAGIC and VERITAS  collaborations. For the 
EBL  we  adopted  the P00 model.  The calculations are performed for a monoenergetic
spectrum of primary gamma-rays with $E_0=100 \ \rm TeV$ and luminosity 
$L_0=10^{45}  \  \rm  erg/s$.

\begin{figure}[t]
\begin{center}
  \scalebox{0.35}{\includegraphics{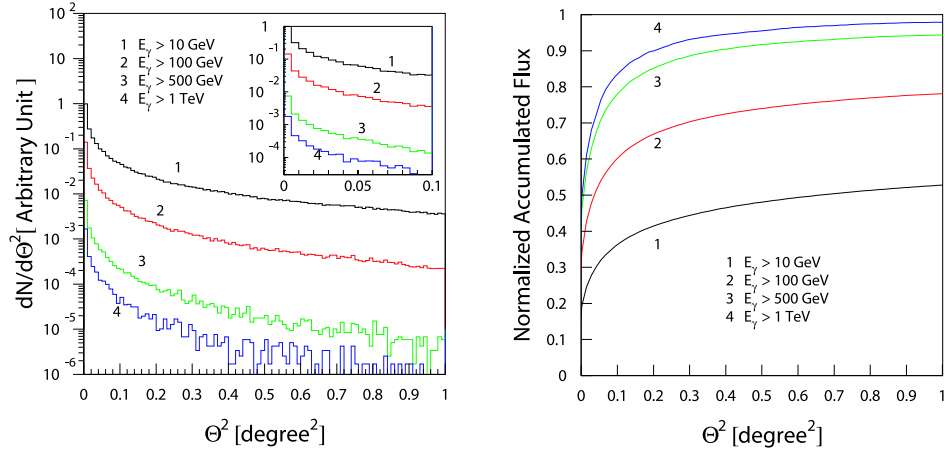}}
  \caption{\small{Left panel: differential 
   angular distributions of the pair halo at $z=0.129$ (in arbitrary units); 
   Right panel: cumulative
   angular distributions  at different energies.}}
   \label{fig:preliminary result3a}
\end{center}
\end{figure}

\begin{figure}[b]
  \begin{center}
   \scalebox{0.8}{\includegraphics{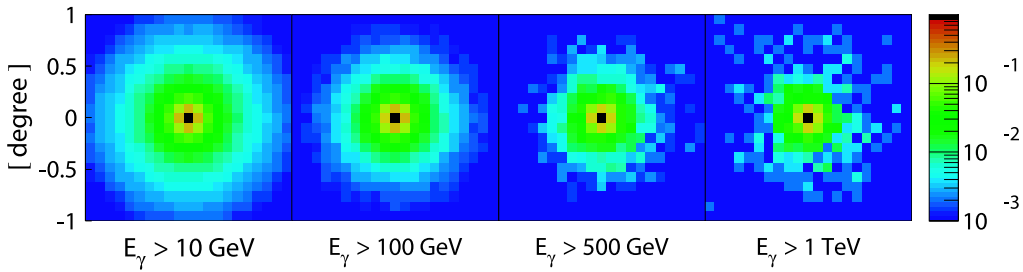}}
   \caption{ \small{Two dimensional map of  a  pair halo  at $z=0.129$ 
   at different  gamma-ray energies.}}
   \label{fig:h1426Halo}
  \end{center}
 \end{figure}

The  differential angular distribution  of gamma-ray flux of the pair halo in 
different energy bands is shown  in  Fig.~\ref{fig:preliminary result3a} (left panel).  
The angular distribution of the  cumulative flux, 
$J_\gamma (\geq \Theta^2)/J_\gamma \rm \rm (all \ angles)$ 
is shown in  Fig.~\ref{fig:preliminary result3a} (right panel).  The initial parameters are the same as in
Fig.~\ref{fig:preliminary result2a}.  As expected at higher energies =
the halo looks more compact than at 
lower energies. This is clearly seen also  in Fig.~\ref{fig:h1426Halo}, where the two-dimensional projections
of pair halos at different energies are  shown.

\begin{figure}[tb]
\begin{center}
   \scalebox{0.35}{\includegraphics{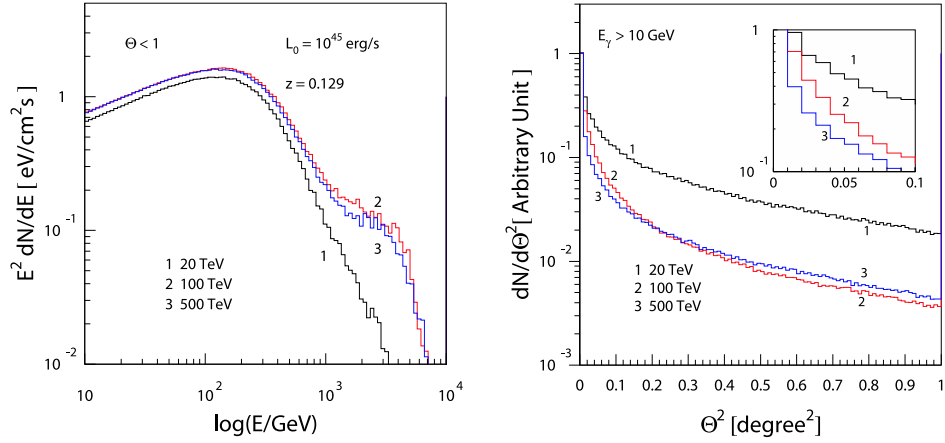}}
   \caption{\small{Radiation of pair halos initiated by monoenergetic gamma-rays with E$_0$ = 20, 100
   and 500 TeV. Left panel:  SEDs;  Right panel:  angular distributions.}}
   \label{fig:compare monoenergetic1}
\end{center}
\end{figure}

\subsection{The impact of the initial gamma-ray spectrum}
 In order to study the impact of the initial gamma-ray spectrum on the basic characteristics 
of pair halos,  we calculated the  angular and spectral distributions of gamma-rays for a halo initiated 
by monoenergetic  gamma-rays. The results are shown in  
Fig.~\ref{fig:compare monoenergetic1} for three different energies of primary gamma-rays: 
E$_0$ = 20, 100, and  500 TeV. One can see that at the intermediate (around the 
maximum of SED) 
and low energies both the energy and angular distributions are quite insensitive to the primary 
energy. The latter becomes important at higher energies, namely between 1 TeV and 
10 TeV.  In particular,  with an increase of $E_0$ the angular distribution becomes sharper, 
especially below $0.1^\circ$.  Also, with an increase of $E_0$,  the energy spectrum
beyond the maximum around a 100-300 GeV  becomes significantly flatter. This tendency continues,
however,   until $E_0 \sim 100$~TeV.  At higher energies of primary gamma-rays,  the results of  
the spectral and angular distributions of the halo  radiation becomes almost insensitive to $E_0$.

\subsubsection{Power law sources}
\begin{figure}[t]
\begin{center}
 \scalebox{0.35}{\includegraphics{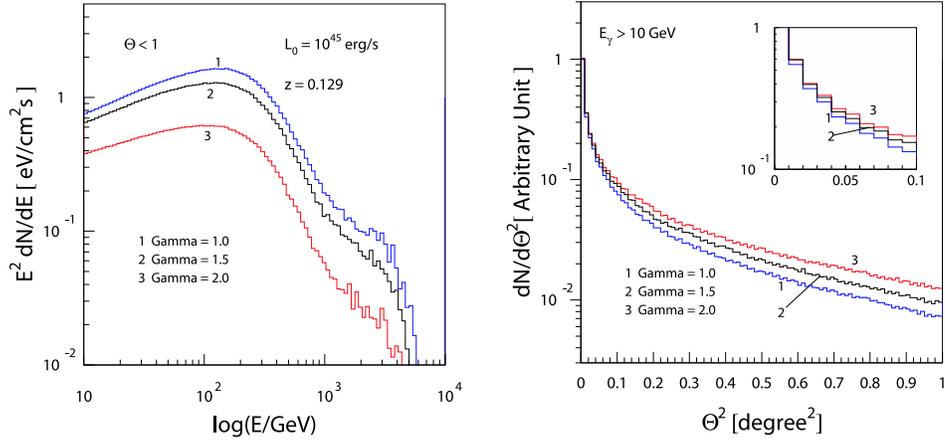}}
 \caption{\small{Pair halos initiated by primary  gamma-rays with 
 power-law spectrum  
 between  $E_{\rm min}$ = 100 GeV and $E_{\rm max}$ = 100 TeV for  
 different photon indices  $\Gamma$=1, 1.5, and 2. 
 Left  Panel -  SED; Right Panel  - angular distribution.}}
 \label{fig:halo Gamma}
\end{center}
\end{figure}

\begin{figure}[t] 
\begin{center}
 \scalebox{0.35}{\includegraphics{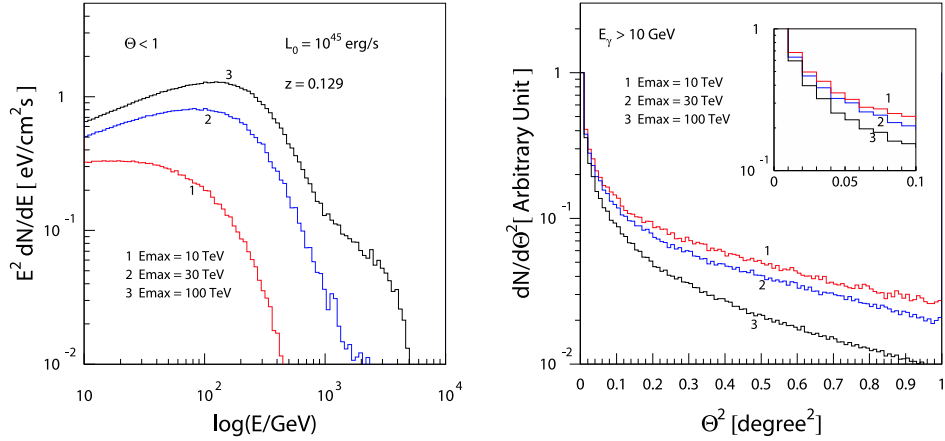}}
 \caption{\small{Pair halos  calculated for a power-law spectrum of primary gamma-rays with  $\Gamma$= 1.5
 and  three  different  $E_{\rm max}$=10 TeV, 30 TeV, and 100 TeV. 
 Left panel- SEDs, Right panel - angular distributions.}}
 \label{fig:halo Emax}
\end{center}
\end{figure}

In  Fig.~\ref{fig:halo Gamma} we show the energy and angular distributions of 
of radiation  of  the pair halo for more realistic, power-law distribution of 
primary gamma-rays, $Q(E) \propto E^{-\Gamma}$ between 100 GeV and 100 TeV, and 
$Q(E)=0$ outside this interval. The results calculated for three different values
of $\Gamma$=1, 1.5, and 2, but for the fixed  total gamma-ray luminosity
$L_0=10^{45} \ \rm erg/s$ are shown in  Fig.~\ref{fig:halo Emax}.  The result are quite similar
which is explained by the fact that for the chosen  hard energy spectra of primary
gamma-ray, the main contributions come  from the energy interval close  to $E_{\rm max}=100$~TeV. 

In Fig.~\ref{fig:halo Emax} we show the case of primary gamma-rays with fixed 
power-law photon index $\Gamma=1.5$, and  for different maximum energies, 
$E_{\rm max}$ = 10, 30, 100 TeV. It is interesting that the pair halo SEDs are more sensitive to 
$E_{ \rm max}$ between 10 to 30 TeV, whereas the angular distributions are more sensitive to 
$E_{ \rm max}$ between 30 to 100 TeV.
 
\subsection{Redshift Dependence}
\begin{figure}[b]
\begin{center}
 \scalebox{0.35}{\includegraphics{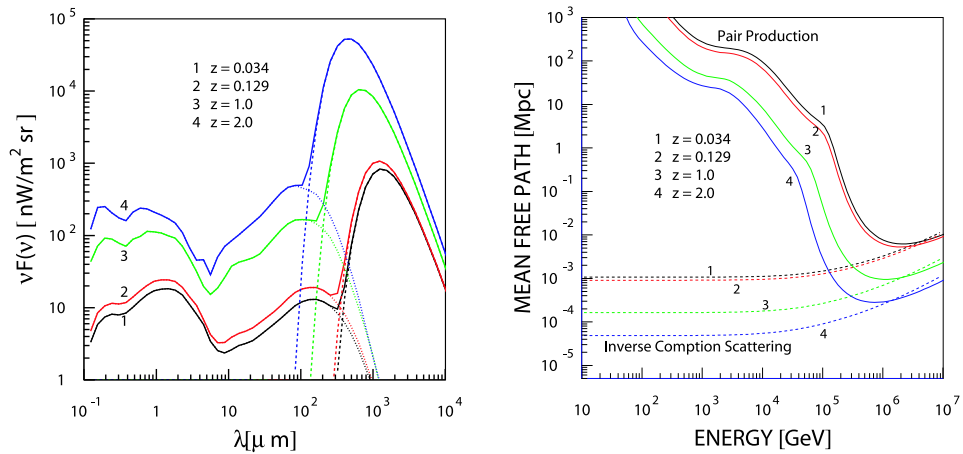}}
 \caption{Left panel: EBL at different
 redshifts; Right panel: mean free paths of  gamma-rays due to 
  pair production (PP) and  electrons due to  the inverse Compton (IC)  
  scattering at different epochs.} 
 \label{fig:background in many redshifts}
\end{center}
\end{figure}

 When the redshift of the source exceeds 0.1 one 
 has to take into account the evolution of EBL in the past. 
 The energy density of the CMB increases with $z$ 
 proportional to  $(1+z)^4$, while  the  peak of distribution 
 shifts  to higher energies  by a  factor of  $(1+z)$  
 (see Fig~\ref{fig:background in many redshifts}).
 The EBL evolution is different from the CMB evolution 
 since both the stellar-light and the dust  
 component of EBL are produced at much later epochs  with $z \lesssim 3$.

\begin{figure}[t]
   \begin{center}
    \scalebox{0.35}{\includegraphics{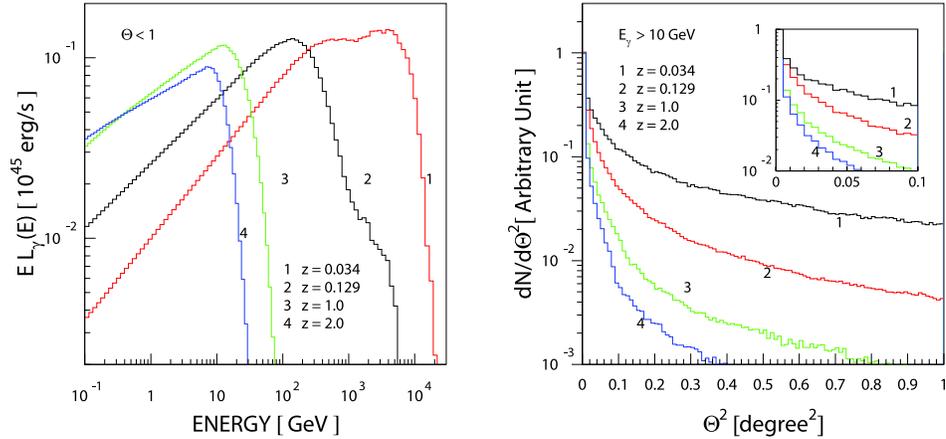}}
    \caption{\small{Spectral luminosities of pair halos  calculated for sources located at 
    different redshifts, $z$ = 0.034, 0.129, 1 and 2, for the opening angle $1^\circ$
    (left panel).  Angular distributions of gamma-rays at energies above 10 GeV (right panel). }} 
    \label{fig:compare redshifts halos}
   \end{center}
  \end{figure}

 In Fig.~\ref{fig:background in many redshifts}  we show
 the evolution of EBL, i.e.  the EBL  flux at different epochs  $z$ 
 in accordance with the P00 model (left panel) .  The corresponding mean 
 free paths  are shown in the right panel of 
 Fig.~\ref{fig:background in many redshifts}. One can see
 that at  the epoch $z=2$, the mean free paths of gamma-rays and 
 electrons in the  intergalactic medium  were an order of magnitude 
 shorter than at the present epoch.  This should obviously result in 
 significant shift of the SED towards low energies and more compact halos
 (in addition to the effects  introduced  by the redshift itself). This is demonstrated 
 in  Fig.~\ref{fig:compare redshifts halos}, where the spectral luminosities and angular distributions 
 of radiation of  pair halos are shown for 
 different redshifts of central sources.  The  total luminosity of primary gamma-rays  
 is assumed $10^{45} \ \rm erg/s$.

\subsection{Dependence on the EBL models}
\begin{figure}[t]
   \begin{center}
    \scalebox{0.35}{\includegraphics{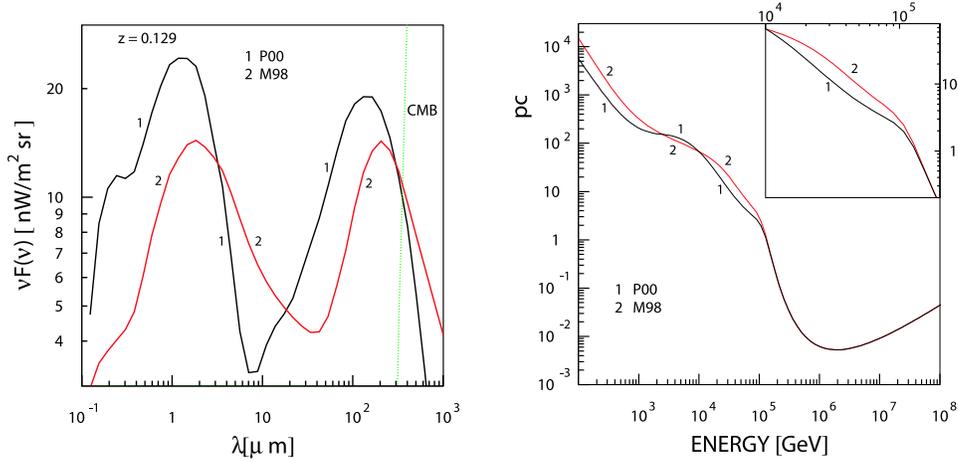}}
    \caption{\small{Left panel: Spectral energy distributions 
    of the P00 and M98 EBL models at the present epoch, $z=0$; 
    Right panel: the mean free paths of gamma-rays for these two models, 
    respectively.}} \label{fig:diff EBLs}
   \end{center}
  \end{figure}

  For energy spectrum of primary gamma-rays extended to $E \leq 10^3$~TeV, 
  the $e^\pm$ pairs are produced mainly through interactions with the EBL photons.
  Therefore, the choice of the EBL model is principal  
  for calculations of characteristics of pair halos. 
  To study the impact of the EBL on the 
  pair halos, we  calculated the parameters of 
  pair halos  for two different EBL models.
  The EBL from P00 is based on semi-analytic calculations. 
  All physical processes, including 
  the formation and evolution of galaxies are included in the
  code. The EBL model M98 is based on empirical assumptions. It adopts
  an EBL flux at the present epoch ($z=0$, 
  and assumes time evolution proportional to  $(1+z)^{3.1}$.
  These two EBL models for $z = 0.129$, as well as the corresponding 
  mean free paths of gamma-rays  are shown in Fig.~\ref{fig:diff EBLs} .

\begin{figure}[b]
  \begin{center}
    \scalebox{0.35}{\includegraphics{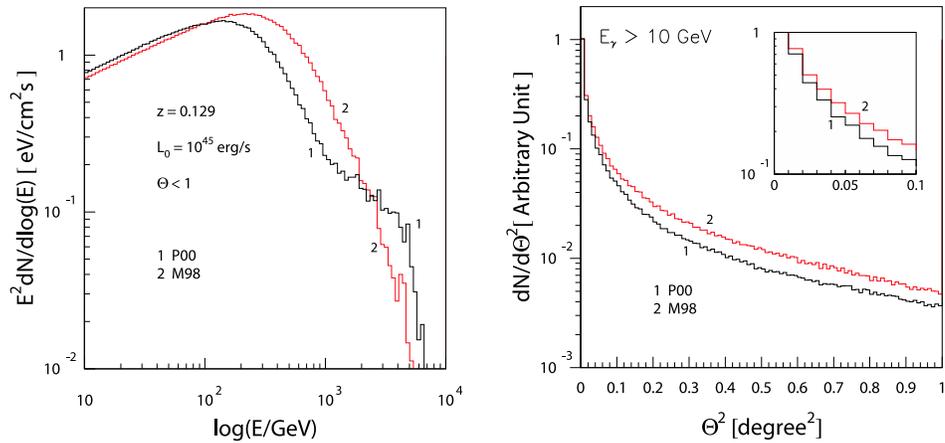}}
    \caption{\small{Left panel: differential angular distributions 
    of radiation of a pair halo at $z=0.129$ calculated for the 
    P00 and M98 models; Right panel: cumulative angular distributions.}} 
    \label{fig:EBL pair halos}
  \end{center}
 \end{figure}

  In Fig.~\ref{fig:EBL pair halos} we show the SEDs and angular distributions of 
  a pair halo located at $z=0.129$ calculated for two EBL models  
  shown in Fig.~\ref{fig:diff EBLs}. One can see that while the SEDs 
  are quite similar at low energies, the energy spectrum corresponding to the 
  P00 model is significantly shifted compared to the SED predicted by the M98
  EBL model. Also, the P00 EBL model predicts sharper angular distribution, especially
  at $\geq 0.1^\circ$. This is explained by the higher EBL flux at mid and far-infrared 
  wavelengths assumed in the P00 model. 

\begin{figure}[t]
  \begin{center}
    \scalebox{0.7}{\includegraphics{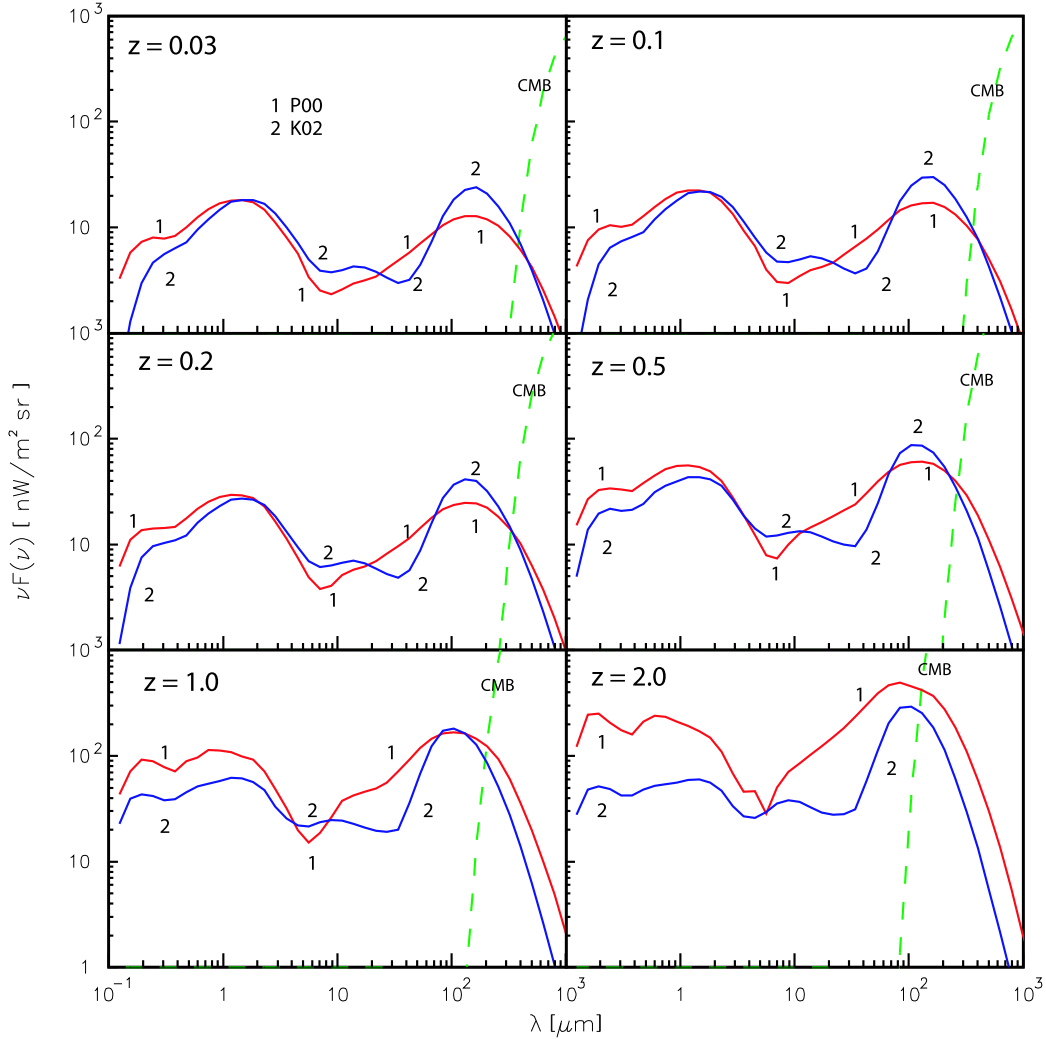}}
    \caption{\small{The EBL fluxes at different redshifts predicted by the 
    P00 and K02 models.}}
    \label{fig:EBL PvsK}
   \end{center}
  \end{figure}

\begin{figure}[t]
  \begin{center}
    \scalebox{0.7}{\includegraphics{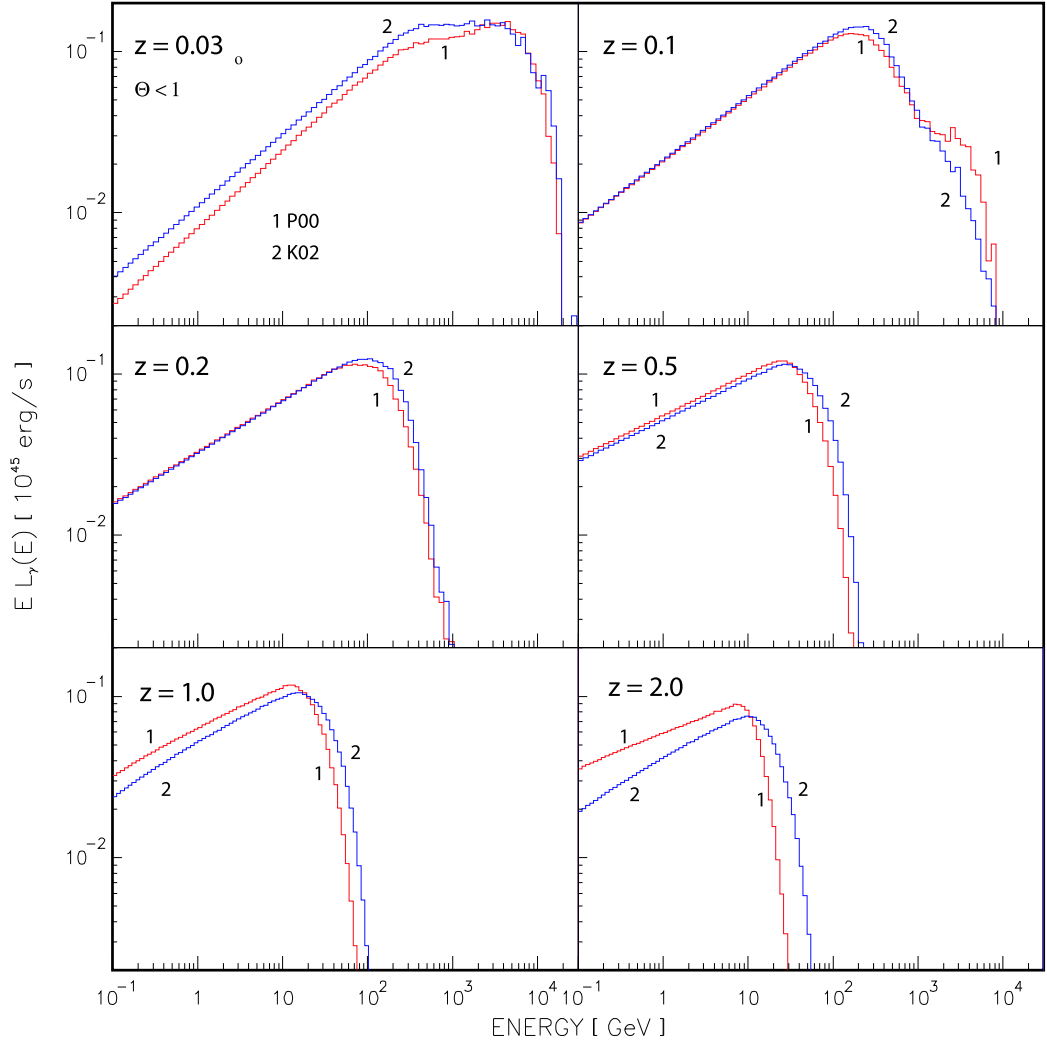}}
    \caption{\small{Spectral luminosity distributions of the pair halos corresponding to each EBL models}}
    \label{fig:SLD PvsK}
   \end{center}
  \end{figure}

 For sources with large redshifts, $z \gg 0.1$, the energy and angular distributions 
 of pair halos depends on the time-evolution of EBL. In fact, this dependence
  is stronger than the dependence on the EBL flux at the present epoch.  
  To demonstrate this effect, in  Fig.~\ref{fig:EBL PvsK} we show the  
  spectral and angular distributions of a halo at $z=2$ calculated 
  for two EBL models - P00 and  K02 \cite{Kneiske_et_al2002}, which are 
  characaterised by significantly different cosmological evolution of the EBL flux.   
 These two models predict relatively similar fluxes at present epoch (or at 
 low redshifts, $z \leq 0.1$). However, at large cosmological epochs, $z\geq 1$,
 the difference of two models becomes as large as an order of magnitude.
 This is seen  in  Fig.~\ref{fig:EBL PvsK}, where the 
 EBL fluxes predicted by the P00 and  K02 model are shown for six 
 different cosmological epochs.  
  The corresponding SEDs and differential angular distributions 
  are shown Fig.~\ref{fig:SLD PvsK} and \ref{fig:DAD PvsK}. One can see that 
  while at low redshifts two EBL models give relatively similar spectral 
  and angular distributions  of gamma-rays, at $z=2$ the difference in predictions 
  of two EBL models becomes 
  significant.  

 \begin{figure}[t]
  \begin{center}
    \scalebox{0.7}{\includegraphics{f17.epsi}}
    \caption{\small{}}
    \label{fig:DAD PvsK}
   \end{center}
  \end{figure}

\section{Summary}
Interactions of high energy gamma-rays with EBL initiate 
electroimagnetic cascades which lead to 
formation of electron-positron halos around
powerful nonthermal extragalactic sources.    
The pair halos contain unique information 
about the flux of EBL and its time-evolution in the past. 
We studied the dependence of 
radiation of pair halos on several key 
parameters, in particular on the 
energy spectrum of the primary gamma-rays, the 
source redshift, and the EBL flux. 
The results presented in this paper can be used 
for search for these giant structures which can 
serve as unique cosmological candles.

\end{document}